\documentclass[twocolumn, aps, superscriptaddress, prb, showpacs,floatfix,longbibliography]{revtex4-1}
\usepackage{dcolumn}
\usepackage{graphicx}
\usepackage{amsmath,amssymb}
\usepackage{bm,color,url}
\usepackage{pict2e,hyperref}

\usepackage{here}

\newcommand{\ket}[1]{|{#1}\rangle}
\newcommand{\bra}[1]{\langle{#1}|}

\begin{document}

\title {Bulk-edge correspondence of Stiefel-Whitney and Euler insulators through the entanglement spectrum and cutting procedure}
\author {Ryo Takahashi}
\affiliation{
Advanced Institute for Materials Research, Tohoku University, Sendai 980-8577, Japan\\
}
\author {Tomoki Ozawa}
\affiliation{
Advanced Institute for Materials Research, Tohoku University, Sendai 980-8577, Japan\\
}

\date{\today}
\begin{abstract}
We propose an unconventional bulk-edge correspondence for two-dimensional Stiefel-Whitney insulators and Euler insulators, which are topological insulators protected by the $PT$ symmetry.
We find that, although the energy spectrum under the open boundary condition is generally gapped, the entanglement spectrum is gapless when the Stiefel-Whitney or Euler class is nonzero. The robustness of the gapless spectrum for Stiefel-Whitney insulator can be understood through an emergent anti-unitary particle-hole symmetry. For the Euler insulators, we propose a conjecture, which is supported by our numerical calculation, that the Euler class is equal to the number of crossing in the entanglement spectrum, taking into account the degree of the crossings. We also discuss that these crossings of the entanglement spectrum are related to the gap closing points in the cutting procedure, which is the energy spectrum as the magnitude of the boundary hopping is varied.

\end{abstract}

\maketitle
\section{Introduction}
Chern insulators are two dimensional energy bands whose Bloch states exhibit non-trivial shape in momentum space, characterized by the topological invariant called the Chern number. Mathematically, Chern number is defined as a consequence of Bloch states being generally \textit{complex} vectors in a certain Hilbert space. For Chern insulators, an important relation known as the bulk-edge correspondence holds, which states that the Chern number is equal to the number of chiral edge modes around the system, which is at the heart of various phenomena such as the integer quantum Hall effect~\cite{PhysRevLett.49.405}.

Recently, a new class of topological phases whose Bloch states are \textit{real} vectors has been proposed~\cite{PhysRevB.92.081201,PhysRevLett.118.056401,PhysRevLett.121.106403,PhysRevX.9.021013,PhysRevLett.125.053601,PhysRevB.102.115135,PhysRevB.103.205303,zhao2022quantum}. Physically such real Bloch states are relevant for $PT$-symmetric systems, where $P$ is the space inversion and $T$ is the time reversal. In the presence of $PT$ symmetry, one can show that the momentum-space Hamiltonian is real, ${H}(\mathbf{k})^* = {H}(\mathbf{k})$, implying that the Bloch wavefunctions can be taken real. 
For such real Bloch bands, topology is not characterized by the Chern number, but by other topological invariants such as the Stiefel-Whitney (SW) classes and Euler classes. SW classes are $\mathbb{Z}_2$ invariants taking either 0 or 1. The first SW class $w_1$, multiplied by $\pi$, corresponds to the quantized Berry phase in one dimension.
In two spatial dimensions, insulators with bands having nonzero second SW class $w_2$ is called the \textit{Stiefel-Whitney insulator}.
Furthermore, if the number of occupied bands is two and $w_1 = 0$ in both directions, the bands can be characterized by the Euler class $\chi$, which can take any integer value~\cite{PhysRevB.96.155105}; the \textit{Euler insulators} refer to situations with $\chi \neq 0$~\cite{PhysRevB.95.235425,PhysRevB.96.155105,PhysRevLett.121.106403,PhysRevX.9.021013,PhysRevB.99.235125}.

Various consequences of having nonzero SW and/or Euler classes have been found in bulk physics.
In three dimensions, non-Abelian braiding of nodal lines was found to be related to the SW classes\cite{bouhon2020non,jiang2021experimental,peng2022phonons}
In two dimensions, nonzero Euler class leads to Wannier obstructions~\cite{PhysRevB.98.085435,PhysRevX.9.021013} and nonzero SW class implies the obstructed atomic insulator~\cite{PhysRevB.99.235125}, and they are relevant for physics of flat bands in the twisted bilayer graphene.
Experimental signature to detect the Euler insulator from quench dynamics has also been proposed~\cite{PhysRevLett.125.053601}. Experiments in acoustic systems\cite{jiang2021experimental} and trapped ions~\cite{,zhao2022quantum} reported detection of nontrivial Euler number through bulk properties.

While the bulk topological properties of $PT$-symmetric insulators have been investigated from various perspectives, little work on the bulk-edge correspondence exists, apart from studies of higher-order topology in such systems~\cite{PhysRevB.99.235125,PhysRevB.106.085129}. One simple reason for the absence of works on bulk-edge correspondence is that the edge spectrum under open boundary condition is generally gapped even when $w_2 \neq 0$ or $\chi \neq 0$, because the $PT$ symmetry is broken on the edges, and thus the bulk-edge correspondence in the usual sense does not hold in these systems.

In this paper, we establish an unconventional bulk-edge correspondence for two-dimensional SW and Euler insulators. We find that, even though the edge spectrum is generally gapped, the \textit{entanglement spectra} of these systems have gapless edge-localized eigenstates which are protected by $PT$ symmetry. 
We also show that these edge-localized entanglement eigenstates are related to spectral flow as one varies the magnitude of the hopping at the boundary, known as the cutting procedure~\cite{PhysRevB.78.045426,song2020twisted,peri2020experimental,PhysRevResearch.2.013300,PhysRevB.101.115120}.

\section{Entanglement spectrum}
The entanglement spectrum is defined as the eigenvalue spectrum of the reduced density matrix, and was proposed as the tool for identifying topological order\cite{PhysRevLett.101.010504,PhysRevB.81.064439}. 
In non-interacting systems, the entanglement spectrum can be determined from the spectrum of the correlation functions\cite{Ingo_Peschel_2003}, 
called single-particle entanglement spectrum, which corresponds to the edge spectrum of spectrally flattened Hamiltonian~\cite{PhysRevLett.104.130502,PhysRevB.82.241102,PhysRevB.83.245132,PhysRevB.84.195103,PhysRevB.85.165120}. The spectrally flatted Hamiltonian is defined in momentum space by
\begin{align}
    H_\mathrm{flat}(\mathbf{k}) &\equiv \frac{\mathbb{I}}{2} - P_\mathrm{occ}(\mathbf{k}), &
    P_\mathrm{occ}(\mathbf{k}) &\equiv \sum_{i:\mathrm{occ}} |u_i (\mathbf{k})\rangle \langle u_i (\mathbf{k})|, 
\end{align}
where $\mathbf{k} = (k_x, k_y)$ is the momentum, $P_\mathrm{occ}(\mathbf{k})$ is the projector to the occupied bands for a given $\mathbf{k}$, $|u_i(\mathbf{k})\rangle$ is the cell-periodic part of the $i$-th Bloch wavefunction, the sum is over the occupied bands, and $\mathbb{I}$ is the identity matrix.
From the $PT$ symmetry, $|u_i(\mathbf{k})\rangle$ can be taken as a real vector, and $P_\mathrm{occ}(\mathbf{k})$ is an $N$-by-$N$ real symmetric matrix, where $N$ is the total number of bands.

The real-space representation of the spectrally flattened Hamiltonian is obtained by Fourier transformation of $H_\mathrm{flat}(\mathbf{k})$. We consider the case where $y$ direction is periodic so that the momentum $k_y$ is taken as a good quantum number, whereas $x$ direction can have an open boundary condition. For a given value of $k_y$, the real space Hamiltonian in $x$ direction with a \textit{periodic boundary condition} is obtained by
\begin{align}
    [H_{\mathrm{flat}}^{x\text{-PBC}}(k_y)]_{x\alpha ,x^\prime \beta} \equiv \frac{1}{L_x}\sum_{k_x}e^{ik_x (x - x^\prime)}\left[H_\mathrm{flat}(\mathbf{k})\right]_{\alpha,\beta},
\end{align}
where $L_x$ is the number of sites in $x$ direction, namely $1 \le x \le L_x$ and the same for $x^\prime$, and we take $L_x$ to be an even number. The sum over $k_x$ is for $k_x = 0, 2\pi \frac{1}{L_x}, 2\pi \frac{2}{L_x}, \cdots, 2\pi \frac{L_x - 1}{L_x}$. In order to impose an open boundary condition, we divide the lattice in two parts; the first part, which we call $A$, is for $1 \le x \le L_x/2$ and the second part, which we call $B$ is for $1 + L_x/2 \le x \le L_x$~\footnote{Although we assume A and B to have the same length in order to derive analytical results, results remain essentially the same even when A and B do not have the same length, as long as both A and B are large enough.}.
We then write $H_{\mathrm{flat}}^{x\text{-PBC}}(k_y)$ in the following block-form:
\begin{align}
    H_{\mathrm{flat}}^{x\text{-PBC}}(k_y) = \begin{pmatrix}
    H_{\mathrm{flat}}^{AA}(k_y) & H_{\mathrm{flat}}^{AB}(k_y) \\ 
    H_{\mathrm{flat}}^{BA}(k_y) & H_{\mathrm{flat}}^{BB}(k_y)
        \end{pmatrix}, \label{eq:xpbc}
\end{align}
where, for example, $H_{\mathrm{flat}}^{AB}(k_y)$ is a matrix obtained by restricting $[H_{\mathrm{flat}}^{x\text{-PBC}}(k_y)]_{x\alpha,x^\prime\beta}$ to $1 \le x \le L_x/2$ and $1 + L_x/2 \le x^\prime \le L_x$.
The flattened Hamiltonian for an open boundary condition in $x$ direction is obtained by only looking at $AA$-part of $H_{\mathrm{flat}}^{x\text{-PBC}}(k_y)$, namely
\begin{align}
H^{x\text{-OBC}}_{\mathrm{flat}}(k_y) \equiv H_{\mathrm{flat}}^{AA}(k_y). \label{eq:xobc}
\end{align}
Since the spectrum of $H^{x\text{-OBC}}_{\mathrm{flat}}(k_y)$ is directly related to the entanglement spectrum of the reduced density matrix of the $A$ part, we refer to the energy spectrum of $H^{x\text{-OBC}}_{\mathrm{flat}}(k_y)$ as the entanglement spectrum.
Interestingly, it was previously known in the context of inversion symmetric topological insulators that the entanglement spectrum can be gapless even if the edge spectrum of the original (not-flattened) Hamiltonian is gapped\cite{PhysRevB.82.241102,PhysRevB.83.245132,PhysRevB.85.165120}. 

The entanglement spectrum, which is bounded between $-1/2$ and $1/2$, is symmetric around $0$, and this is due to the emergent antisymmetry, as we shall explain now.
For a fixed $k_y$, the flattened Hamiltonian with open boundary condition $H^{x\text{-OBC}}_{\mathrm{flat}}(k_y)$ has $\frac{L_x}{2}N$ eigenvalues and eigenstates. Some eigenvalues are exactly at $\pm1/2$, but some others distribute between $-1/2$ and 1/2. We collect all eigenstates with eigenvalues not equal to $\pm1/2$ as $|\psi_i\rangle$: $H^{x\text{-OBC}}_{\mathrm{flat}}(k_y) |\psi_i \rangle = \epsilon_i |\psi_i \rangle$, $\epsilon_i \neq \pm1/2$.

Then, as we show in the Appendix \ref{Appendix A}, the operator $\Xi$ defined by
\begin{align}
    \Xi &\equiv \sum_i |\phi_i \rangle\langle \psi_i|,
    \label{Xi_ep_psi}
    \\
    |\phi_i \rangle&\equiv-\frac{H_{\mathrm{flat}}^{BA}(k_y)}{\sqrt{ 1/4 - \epsilon_i^2}}|\psi_i \rangle, 
\end{align}serves as an antisymmetry for states with $\epsilon_i \neq \pm 1/2$. Namely, $H_\mathrm{flat}(k_y) \Xi |\psi_i\rangle = - \epsilon_i\Xi |\psi_i\rangle$. We note that $|\phi_i \rangle=\Xi |\psi_i\rangle$ is normalized as shown in the Appendix \ref{Appendix A}. Generally, the operator $\Xi$ changes the side of edge states. If the state $|\psi_i\rangle$ is localized on the right edge, $\Xi |\psi_i \rangle$ is localized on the left edge. Existence of such an antisymmetry $\Xi$ is a general property of flattened Hamiltonians.

In the presence of the $PT$ symmetry, since $PT$ operation switches the side of edge states but does not change the energy, one can further construct a state $(PT)\Xi |\psi_i \rangle$, which has the energy $ - \epsilon_i$ and localized at the same edge as $|\psi_i \rangle$. As shown in the Appendix \ref{Appendix A}, $(PT)\Xi$ is an antiunitary antisymmetry which squares to one in the subspace spanned by $|\psi_i\rangle$. Namely, we can regard $((PT)\Xi)^2 = +1$ as long as we only consider states with $\epsilon_i \neq \pm 1/2$. This particle-hole like symmetry will play a key role in the robustness of linear crossing of entanglement spectrum as we see below.

\section{Stiefel-Whitney insulator}
Now we study entanglement spectrum of Stiefel-Whitney insulators and show that there is nontrivial linear crossing protected by the $PT$ symmetry.
As a concrete example, we consider the following model\cite{PhysRevLett.121.106403}:
\begin{align}
H_{\text{SW}}(\bm{k})
&=\sin k_x\sigma_x+\sin k_y \tau_y\sigma_y +(1-\cos k_x-\cos k_y)\sigma_z
\notag \\
&\quad +\frac{\tau_z\sigma_z}{2}
+\frac{\tau_x}{4}. 
\end{align}
Here, $\tau_i$ and $\sigma_i$ ($i=x,y,z$) are Pauli matrices. This is a four-band model ($N=4$), and we assume that the lower two bands are occupied. It is known that the model has $w_2 = 1$, which can be confirmed by calculating the Wilson loop spectrum as demonstrated in Appendix \ref{Appendix B}; we note in this case the Euler class can also be defined and $\chi=1$.

\begin{figure}[t]
\centerline{\includegraphics[width=8.5cm,clip]{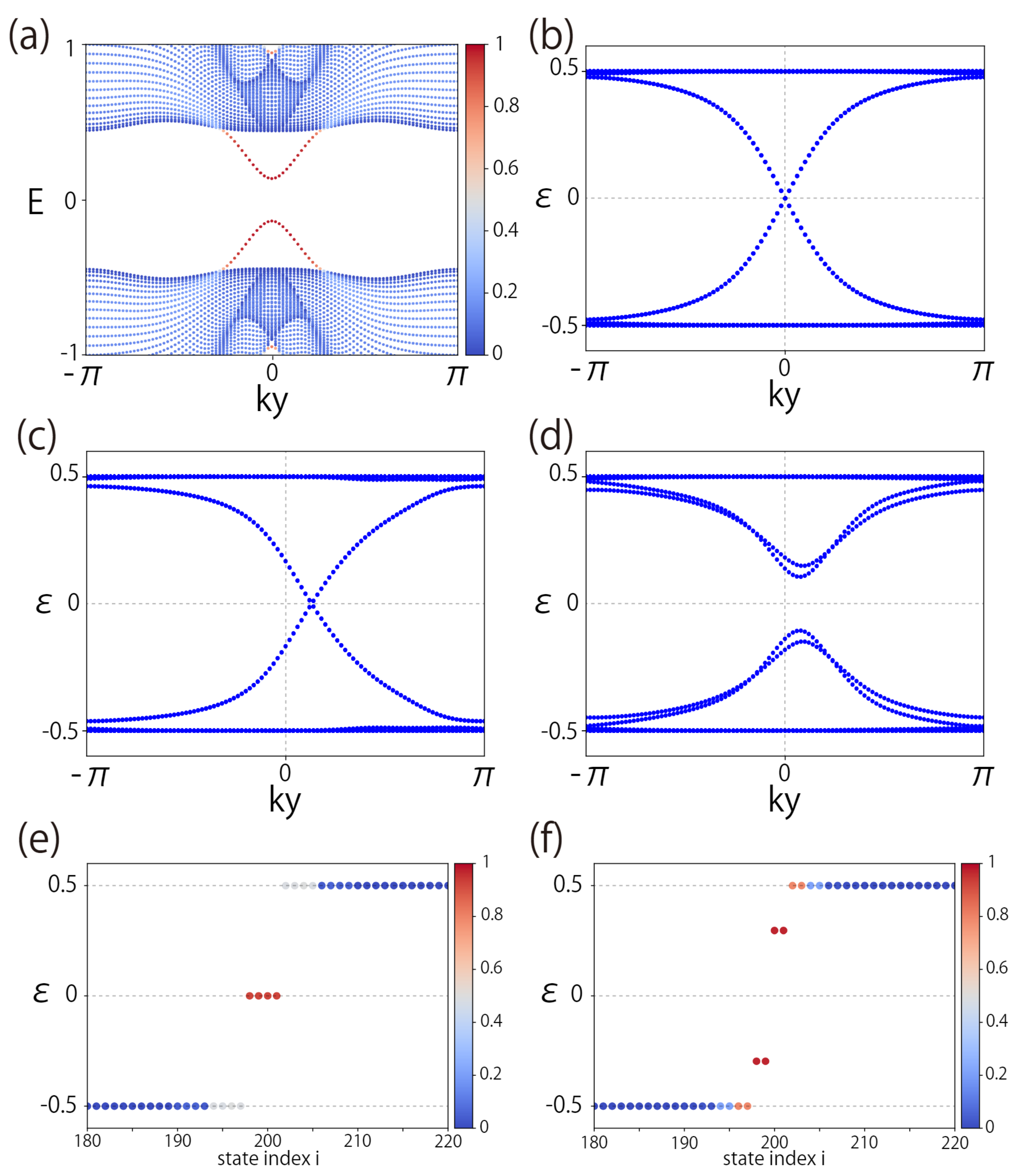}}
  \caption{Numerical calculation for Stiefel-Whitney insulator model and its perturbation. (a) Edge spectrum and (b) entanglement spectrum for $H_{\text{SW}}$. (c,d) Entanglement spectrum with (c) $PT$-preserving mass term [$-M_1\tau_y\sigma_y$] and (d) $PT$-breaking mass term [$M_2(-\tau_y\sigma_y+\tau_x\sigma_x+\tau_y-\sigma_y)$]. Here, $(M_1,M_2)=(0.8,0.4)$. (e,f) Eigenvalue distribution at $k_y=0$ in (b) and (c). Colors in (a,e,f) represent the squared amplitudes of the states in the unit cells on either end. While the edge spectrum of the original (not flattened) Hamiltonian [(a)] is gapped, the entanglement spectrum [(b,c)] exhibits a gapless spectrum, and it becomes gapped under $PT$-breaking term [(d)].
  }
    \label{SW_OBC_ES}
\end{figure}

In Fig.~\ref{SW_OBC_ES}(a), we show the edge spectrum of the origianl (not flattened) Hamiltonian as a function of $k_y$. In this case, the spectrum is clearly gapped; gapless edge spectrum characteristic of conventional topological insulators is absent.
On the contrary, the entanglement spectrum, plotted in Fig.~\ref{SW_OBC_ES}(b), is gapless and show linear crossing at $\epsilon=0$. Here, the entanglement spectrum is symmetric around $\epsilon = 0$ because of the antisymmetry $\Xi$ defined in Eq.~(\ref{Xi_ep_psi}).
Even if we add more terms to the Hamiltonian preserving the $PT$ symmetry, the linear crossing of the entanglement spectrum does not gap out, as exemplified in Fig.~\ref{SW_OBC_ES}(c)]. If we add $PT$-breaking terms to the Hamiltonian, $w_2$ and $\chi$ are no longer well-defined, and the entanglement spectrum can become gapped, as shown in Fig.~\ref{SW_OBC_ES}(d). 

This linear crossing in the entanglement spectrum is the bulk-edge correspondence unique to the SW insulators. In fact, such a linear crossing of the entanglement spectrum was already experimentally observed in trapped ions~\cite{zhao2022quantum}. We now show that this linear crossing is indeed robust against $PT$-symmetric perturbations.
The linear crossing of the entanglement spectrum is the consequence of the anti-unitary particle-hole symmetry $(PT)\Xi$. To see how this symmetry protects the linear crossing, we consider an effective model around the crossing describing edge states on one side. (Note that, while acting with $\Xi$ changes the side of the edge states, $(PT)\Xi$ keeps the side.)
First, we rearrange the basis appropriately so that $(PT)\Xi=K$, where $K$ is the complex conjugation operation. (We explain in Appendix \ref{Appendix C} how one can find such a basis.)
In this basis, since $(PT)\Xi=K$ must anti-commute with the effective Hamiltonian, the effective Hamiltonian must take the following form
\begin{align}
    H_\mathrm{eff}(k_y) = c k_y \sigma_y,
\end{align}
with some nonzero constant $c$. Here, without loss of generality, we took the crossing point to be at $k_y = 0$.
Any perturbation one can add to the system keeping the $PT$ symmetry should also anti-commute with $(PT)\Xi=K$ and thus the only form of the allowed mass term is $m \sigma_y$, and thus the resulting effective Hamiltonian $H_\mathrm{eff}(k_y) = (c k_y + m)\sigma_y$ is still gapless but the position of the gapless point is shifted to $k_y = -m/c$. 

We can generalize the above discussion of $2\times2$ effective model to include more states using the Pfaffian. Taking $(PT)\Xi=K$ as before, from the same argument, the effective Hamiltonian $H_\mathrm{eff}(k_y)$ describing modes localized on one edge should be a skew-symmetric matrix. Then, we can construct a real skew-symmetric matrix $T_{k_y}$ by
\begin{align}
T_{k_y}\equiv iH_{\text{eff}}(k_y).
\end{align}
Since $T_{k_y}$ is a real skew-symmetric matrix, for each momentum, the Pfaffian $\mathrm{Pf}(T_{k_y})$ can be defined. Taking the linear crossing of the entanglement to be at $k_y = 0$, we take two momenta $k_0 < 0$ and $k_1 > 0$, and consider their respective Pfaffians $\mathrm{Pf}(T_{k_0})$ and $\mathrm{Pf}(T_{k_1})$. Even though the overall sign of a Pfaffian is ambiguous, the sign change of the Pfaffian between $k_0$ and $k_1$ is a valid invariant of the linear crossing, assuming that the effective Hamiltonian is defined continuously in the region $k_0 \le k_y \le k_1$ as discussed in [\onlinecite{carey2019spectral}]. To see that the sign change of the Pfaffian is related to the protection of the linear crossing, first we note that the effective Hamiltonian can be written in block-diagonal form, and then the Pfaffian can be simply calculated as product of $\epsilon_i$'s:
\begin{align}
\text{Pf}(\text{diag}[i\epsilon_1\sigma_y,\cdots,i\epsilon_{N_0}\sigma_y])=\epsilon_1\times\cdots\times\epsilon_{N_0}. \label{eq:pfaffian}
\end{align}
Thus the sign change of the Pfaffians is the same as the sign change of $\prod_{i}\epsilon_i(k_y)$ between $k_0$ and $k_1$. Bands that do not close the gap between $k_0$ and $k_1$ do not contribute to the sign change, and thus the change of the sign of the Pfaffian implies that two eigenvalues with opposite values cross somewhere in $k_0 < k_y < k_1$.
Moreover, even if a perturbation that hybridizes different bands is added, the sign change is preserved as long as the spectral gap at $k_0$ and $k_1$ remains open. Thus, the linear degeneracy cannot open a gap with a perturbation which keeps the $PT$ symmetry. 

Finally, we comment on the model independence of our results. The discussion given in Ref.~\cite{PhysRevB.102.115135} shows that the homotopic classification of the $PT$-symmetric gapped Hamiltonians coincide with the classification based on the SW and Euler classes. This implies that if two Hamiltonians have the same SW class, they can be continuously transformed to each other, up to addition of trivial bands, without closing the gap. Therefore, since we have shown that a crossing in the entanglement spectrum remains robust under continuous deformations, we can conclude that a linear crossing of entanglement spectrum exists for any two-dimensional $PT$-symmetric models with nontrivial SW class.

\section{Euler insulator}
Next, we consider the Euler insulator with $\chi=2$. We find that there is either one quadratic crossing or two linear crossings in the entanglement spectrum.  
We first consider the following three band model
\cite{PhysRevLett.125.053601,PhysRevB.103.205303,zhao2022quantum}:
\begin{align}
&H_{\text{Euler}}
=
\begin{pmatrix}
d_1^2 & d_1d_2 & d_1d_3 \\
d_1d_2 & d_2^2 & d_2d_3 \\
d_1d_3 & d_2d_3 & d_3^2
\end{pmatrix}
+c_1
\begin{pmatrix}
d_2^2 & -d_1d_2 & 0 \\
-d_1d_2 & d_1^2 & 0 \\
0 & 0 & 0
\end{pmatrix},
\\
&(d_1,d_2,d_3)
=(\sin k_x,\sin k_y, m-\cos k_x-\cos k_y),
\end{align}
where $m$, $c_1$ are real parameters, which we take to be $(m,c_1)=(1.2,0.1)$. 

\begin{figure}[t]
\centerline{\includegraphics[width=8.5cm,clip]{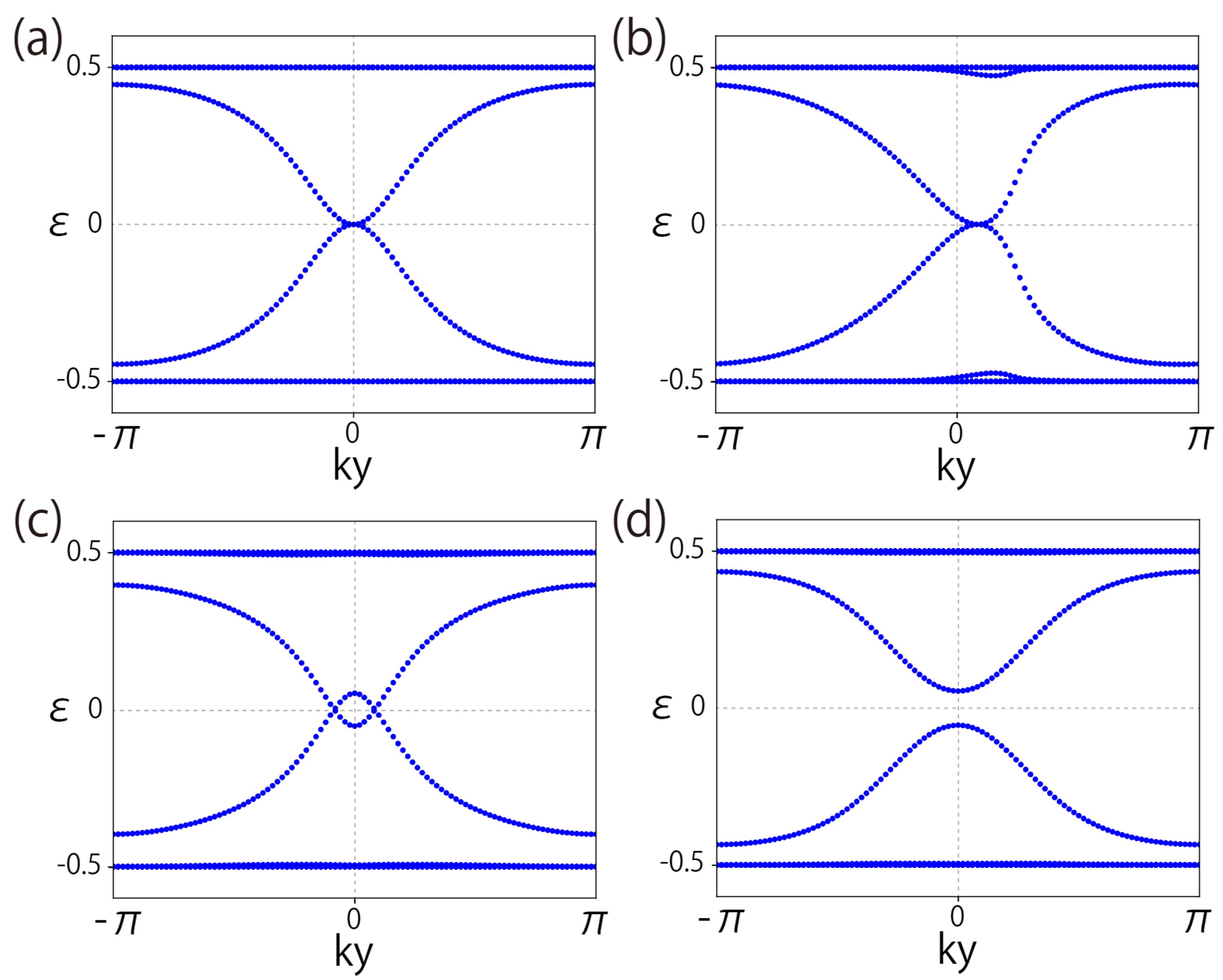}}
  \caption{Entanglement spectrum for (a) $H_{\text{Euler}}$, (b) $H_{\text{Euler}}-\frac{\sin k_x}{2}\bigl(\begin{smallmatrix} 0&1&0 \\ 1&0&0 \\ 0&0&0 \end{smallmatrix}\bigr)$, and that for $H_{\text{Euler+NI}}$ with (c) $V_0=1.5$ and (d) $V_0=-0.5$. These systems have (a-c) two and (d) three occupied bands, and have topological numbers (a-c) $\chi=2$ and (d) $w_2=0$ (mod $2$), respectively.  
  }
    \label{EZ+NI_ES}
\end{figure}
Considering the lower two bands to be occupied, the Euler class can be defined, and it is $\chi = 2$.
(We note that SW class is given by $w_2 = \chi \ \text{mod } 2 = 0$.)
We plot the entanglement spectrum in Fig.~\ref{EZ+NI_ES}(a); the entanglement spectrum shows a quadratic degeneracy. By adding $PT$-symmetric perturbations to the Hamiltonian, the position of the quadratic degeneracy can shift, but we have numerically confirmed that a gap never opens unless the bulk band gap is closed (Fig.~\ref{EZ+NI_ES}(b)).

We can further understand the nature of this crossing of the entanglement spectrum by coupling an additional trivial band to the model:
\begin{align}
H_{\text{Euler+NI}}(\bm{k})=
\begin{pmatrix}
H_{\text{Euler}}&\Phi^{\dagger}\\
\Phi&V_0
\end{pmatrix},
\\
\Phi=\Delta
\begin{pmatrix}
\sin(k_x+\frac{\pi}{4}),& 0,& \frac{1}{2}\sin(k_x-\frac{\pi}{4})
\end{pmatrix},
\end{align}
where $\Phi$ is the coupling term and we set $\Delta=1.0$. 
The original $H_{\text{Euler}}$ has a bandgap around $E=0.2$ to 0.65.
By turning on $\Delta$, one can add a band either in the occupied or unoccupied band by adjusting $V_0$ to be below or above the bandgap.
First, we consider adding a trivial band to the occupied bands by setting $V_0=-0.5$. 
We consider the lower three bands, instead of the two bands, to be occupied. Since defining the Euler class requires the number of occupied bands to be two, the Euler class in this new model is no longer defined. The SW class can be still defined, and since the trivial band considered here can be added adiabatically without closing the gap, SW class remains to be $w_2 = 0$.
As a band is added, the entanglement spectrum gaps out, as shown in Fig.~\ref{EZ+NI_ES}(d), which suggests that the quadratic band crossing of the Euler insulator is protected by the Euler class $\chi = 2$. 

Next, we consider adding a trivial band to unoccupied bands by setting $V_0=1.5$. 
 Considering two lower bands to be occupied, the Euler class remains to be $\chi = 2$.
In Fig.~\ref{EZ+NI_ES}(c), we show the entanglement spectrum. We see that the quadratic crossing is now split into two linear crossings. We have numerically checked that, adding any perturbations up to next-nearest neighbor hoppings, the entanglement spectrum is always gapless as long as the bulk energy gap is kept open, and either one quadratic crossing or two linear crossings exist.

From our numerical calculation, we conjecture that the Euler class equals to the total number of band crossings in the entanglement spectrum, taking into account the order of crossings:
\begin{align}
|\chi|=\sum_{p\in\mathbb{N}}p\cdot N_p, \label{eq:conjecture}
\end{align}
where $N_p$ is the number of band crossings with degree $p$ in the entanglement spectrum; in particular, $p=1$ corresponds to linear crossings and $p=2$ to quadratic crossings.
To further support our conjecture, we analyzed three more models of Euler insulators in Appendix \ref{Appendix D}. We have confirmed that Eq.~(\ref{eq:conjecture}) is indeed obeyed by all the models we analyzed.

We note that the robustness of linear crossings in Euler insulators, such the ones in Fig.~\ref{EZ+NI_ES}(c) can be understood from the same reason as in the SW insulator. However, the reasoning we used for the SW insulator does not explain why the quadratic crossing, around which the Pfaffian (\ref{eq:pfaffian}) does not change the sign, is robust against $PT$-symmetric perturbations; further investigation is required to fully understand the origin of this robustness.

\section{Cutting procedure}
Finally, we discuss that the crossing of the entanglement spectrum of SW and Euler insulators is related to the degeneracy of edge spectrum upon one varies the magnitude of the hopping at the boundary, a process known as the cutting procedure~\cite{PhysRevB.78.045426,song2020twisted,peri2020experimental,PhysRevResearch.2.013300,PhysRevB.101.115120}. 
In the cutting procedure, we continuously connect the original (non-flattened) Hamiltonian with periodic boundary conditions $H_\mathrm{PBC}$ and the one with open boundary conditions $H_\mathrm{OBC}$ by one parameter $\lambda$: 
\begin{align}
H_{\lambda}=\lambda H_{\text{PBC}}+(1-\lambda)H_{\text{OBC}}.
\end{align}
(We note that $H_\mathrm{PBC}$ and $H_\mathrm{OBC}$ differ just by the boundary hopping, and in particular share the same size.)
The cutting procedure has been used to analyze the bulk-boundary correspondence of higher-order topological insulators and fragile topological insulators, which do not have gapless edge states in the simple open boundary condition\cite{song2020twisted,peri2020experimental,PhysRevResearch.2.013300,PhysRevB.101.115120}. 
We numerically found that SW and Euler insulators exhibit non-trivial spectral flow with $2w_2$ or $2\chi$ band crossing points. 

\begin{figure}[t]
\centerline{\includegraphics[width=8.5cm,clip]{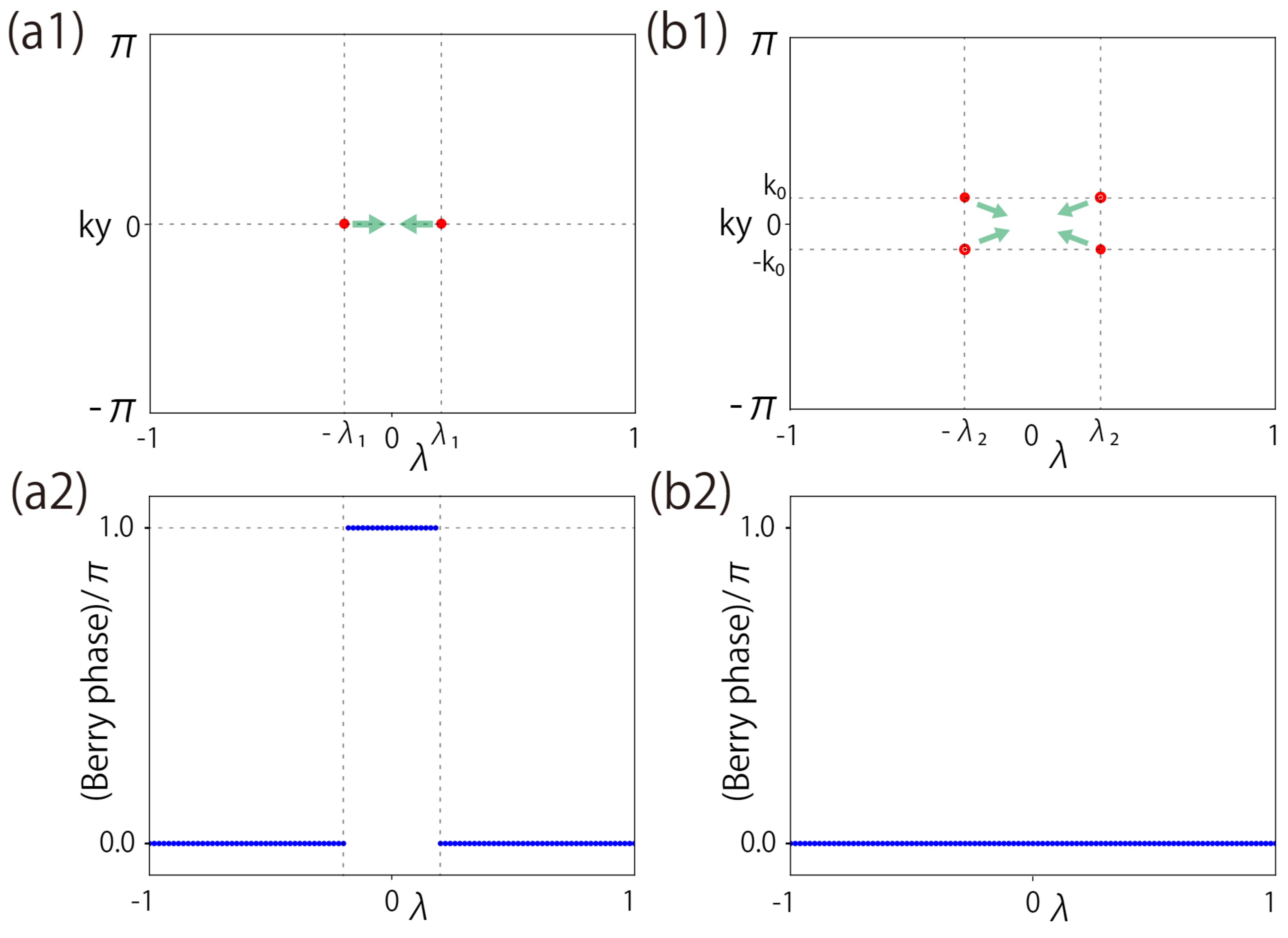}}
  \caption{Spectral flow of (a) Stiefel-Whitney insulator and (b) Euler insulator with cutting procedure. Position of the gap closing points are (a1) $(\pm\lambda_1,0)$ and (b1) $(\pm k_0,\pm\lambda_2)$, where $\lambda_1=0.2$, $\lambda_2=0.28$ and $k_0=0.44$. As the bulk Hamiltonian is flattened, these degeneracy points approach $(\lambda,k_y)=(0,0)$, eventually reproducing the degeneracy point of the entanglement spectrum. (a2,b2) The Berry phase distribution of (a1) and (b1).
  }
    \label{TBC_SF}
\end{figure}
The spectral flow of the SW insulator $H_{\text{SW}}$ is shown in Fig.~\ref{TBC_SF}(a), which have $2w_2=2$ gap closing points. 
The energy dispersion around the gap closing points are linear in both $\lambda$ and $k_y$.
In the presence of the $PT$ symmetry, one can show that the Berry phase calculated along any closed path in $\lambda$-$k_y$ space is quantized either to 0 or $\pi$~\cite{PhysRevB.93.205132}. 
Moreover, the Berry phase around each of these gap closing points is quantized to $\pi$, and thus they are protected because one cannot continuously change the Berry phase from $\pi$ to 0. as shown in Fig.~\ref{TBC_SF}(a2)
As a result, the Berry phase in the $k_y$ direction changes by $\pi$ across the degeneracy point as shown in Fig.~\ref{TBC_SF}(a2).
Here, the Berry phase in the $k_y$ direction is defined as follows:
\begin{align}
\theta^{\text{Berry}}_y(\lambda)
=\text{Im}\Bigg{[}
\int_0^{2\pi}dk_y \text{Tr}[A_y(k_y,\lambda)]
\Bigg{]},
\\
[A_y(k_y,\lambda)]_{ij}
=\Big{\langle} u^i(k_y,\lambda)\Big{|}
\frac{\partial}{\partial k_y}u^j(k_y,\lambda)\Big{\rangle},
\end{align}
where $|u^i(k_y,\lambda)\rangle$ is the $i$-th eigenstate of $H_{\lambda}(k_y)$ below the gap, and $A_y (k_y, \lambda)$ is the Berry connection matrix. We note that the size of $A_y (k_y, \lambda)$ is $2N_x \times 2N_x$, where $N_x$ is the number of unit cells along $x$ direction, and 2 accounts for the number of occupied bulk bands.

As one flattens the Hamiltonian, the gap closing points move and merge at $\lambda = 0$ line~\footnote{Here, we need to consider a parametrized Hamiltonian of the form $H_{\mathrm{flat},\lambda} = \lambda H_{\mathrm{flat}}^{x\text{-PBC}} + (1-\lambda) H_{\mathrm{flat}}^{x\text{-OBC}}$. However, our construction of the flattened Hamiltonian around Eqs.~(\ref{eq:xpbc}-\ref{eq:xobc}) implies that the size of $H_{\mathrm{flat}}^{x\text{-PBC}}$ is twice larger than that of $H_{\mathrm{flat}}^{x\text{-OBC}}$. In order to correctly define $H_{\mathrm{flat},\lambda}$, we calculate  $H_{\mathrm{flat}}^{x\text{-PBC}}$ with half the system size so that the size of $H_{\mathrm{flat}}^{x\text{-PBC}}$ and $H_{\mathrm{flat}}^{x\text{-OBC}}$ match.}.
The limit of perfectly flattened Hamiltonian corresponds to the gap closing point at $\lambda = 0$; thus the edge crossing of the entanglement Hamiltonian can be understood as a limit of the gap closing points in the spectral flow of the cutting procedure.

We also calculate the spectral flow of the Euler insulator $H_{\text{Euler}}$ in Fig.~\ref{TBC_SF}(b1), which have $2\chi=4$ gap closing points, which are all linear crossings in both $\lambda$ and $k_y$ directions. 
In this case, the Berry phase $\theta^{\text{Berry}}_y(\lambda)$ changes by $2\pi\equiv0$ (mod $2\pi$) at $\lambda=\lambda_2$, because there are two degenerate points at $\lambda=\lambda_2$, and for each degenerate point the Berry phase should change by $\pi$. The same holds for $\lambda = -\lambda_2$. The Berry phase thus remains unchanged as one crosses $\lambda=\pm \lambda_2$ as shown in Fig.~\ref{TBC_SF}(b2). 
We numerically find that these band crossing points are robust as long as the Euler class $\chi = 2$ is unchanged.
We have also checked that, by flattening the Hamiltonian, these gap closing points merge along the line $\lambda = 0$, recovering the result of the entanglement spectrum.

\section{Conclusion and discussion}
In this paper, we found an unconventional bulk-edge correspondence for two-dimensional lattice models with $PT$ symmetry. For the insulator characterized by the second SW class, there is either one or zero linear edge crossing in the entanglement spectrum depending on either $w_2 = 1$ or $w_2 = 0$. This edge crossing is protected by the emergent anti-unitary particle-hole symmetry in the entanglement spectrum. For the Euler insulator, we proposed a conjecture that the Euler class equals to the sum of the number of crossings in the entanglement spectrum taking degree into account, which is supported by our numerics. 
We have also found that the crossings of entanglement spectrum have precursors in the gap closing points of the spectral flow of the cutting procedure. 

In recent years, there is an increasing interest in experimentally detecting the entanglement spectrum using various engineered quantum materials such as ultracold atomic gases\cite{PhysRevX.6.041033,dalmonte2018quantum,PhysRevResearch.3.013112}; experimental measurement of the entanglement spectrum was already demonstrated using the IBM quantum computers\cite{PhysRevLett.121.086808}. In trapped ions, the entanglement spectrum of Euler insulators was also experimentally reconstructed from the measurement of bulk eigenstates~\cite{zhao2022quantum}. The parameter($\lambda$)-dependent Hamiltonian we introduced for the cutting procedure is also experimentally accessible in metamaterials~\cite{peri2020experimental}. In light of these rapid experimental developments, the features of the bulk-edge correspondence of SW and Euler insulators discussed in this paper should be experimentally verifiable in a near future.

Our results show that the $PT$-symmetric topological models have unconventional bulk-edge correspondences. 
In three-dimensional bulk spectra, examples where the charge of the Weyl point and the degree of the band crossing are related are known~\cite{PhysRevB.102.125148}, but, to our knowledge we are not aware of any other systems where the degree of edge crossing should be taken into account upon considering the bulk-edge correspondence.
While the SW insulator case can be clearly understood, the Euler insulator requires further investigation; more rigorous proof of the bulk-edge correspondence of the Euler insulator is left for future works. It is also of interest to extend the entanglement-spectrum analysis of the bulk-edge correspondence to other symmetry classes in the AZ+$\mathcal{I}$ classification~\cite{PhysRevB.96.155105}, which is an extension of the AZ classification that includes $PT$ symmetry.
 We expect that similar analysis of the bulk-edge correspondence through the entanglement spectrum and/or cutting procedure will be of use for further analysis of fragile and delicate topological insulators.

\begin{acknowledgments}
We are grateful to helpful discussion with Tom\'{a}\v{s} Bzdu\v{s}ek and Shin Hayashi. This work is supported by JSPS KAKENHI Grant No. JP20H01845, JST PRESTO Grant No. JPMJPR19L2, and JST CREST Grant No.JPMJCR19T1.
\end{acknowledgments}

\appendix{}

\section{Emergent particle-hole symmetry}\label{Appendix A}
To understand the emergent particle-hole symmetry in the flattened Hamiltonian, it is convenient to introduce the following \textit{correlation matrix} $C_X$\cite{PhysRevB.83.245132,PhysRevB.85.165120,zhao2022quantum}:
\begin{align}
    C_X (k_y) := \frac{\mathbb{I}}{2} - H_\text{flat}^{x\text{-PBC}}(k_y).
\end{align}
Each element is defined by
\begin{align}
    \left[ C_X (k_y)\right]_{x\alpha, x^\prime \beta} = \frac{1}{L_x}\sum_{k_x} e^{ik_x (x-x^\prime)}\left[ P_\mathrm{occ}(\mathbf{k})\right]_{\alpha, \beta}.
\end{align}
The reason we want to introduce $C_X (k_y)$ is that this is a \textit{projector}, that is, $C_X(k_y)^2 = C_X(k_y)$ holds.
We define four blocks of $C_X (k_y)$ by
\begin{align}
C_X (k_y) = 
    \begin{pmatrix}
C_{AA}(k_y) & C_{AB}(k_y) \\
C_{BA}(k_y) & C_{BB}(k_y)
\end{pmatrix}.
\end{align}
Here, as in the main text, we divided $x$ direction in two parts $A$ and $B$.
Since $C_X$ is a Hermitian matrix, we see that $C_{AB} = C_{BA}^\dagger$.
Assuming that $A$ and $B$ regions have the same number of lattice sites, one can also generally show that $C_{AA}=C_{BB}$ and $C_{AB}=C_{BA}$. 
The spectrally flattened Hamiltonian in the open boundary condition and the upper left block of the correlation matrix are related by
\begin{align}
H^{x\text{-OBC}}_{\text{flat}}(k_y)
=\frac{\mathbb{I}}{2}-C_{AA}(k_y).
\label{H_flat_CL}
\end{align}
Therefore, they share the same eigenstates. 

We first show that $PT$ symmetry relates two eigenstates of $C_{AA}$ with the same eigenvalue.
As in the main text, we assume that the $PT$ operation is represented by complex conjugation $K$ in the Bloch basis. Then, the projection to occupied states is invariant by complex conjugation:
\begin{align}
P^{*}_{\text{occ}}(\mathbf{k})
=P_{\text{occ}}(\mathbf{k}). 
\end{align}
This leads to the following symmetry for $C_{X}$:
\begin{align}
[C_X(k_y)]^{*}_{x\alpha,x'\beta}
&=\frac{1}{L_x}\sum_{k_x}e^{-ik_x(x-x')}[P_{\text{occ}}(\mathbf{k})]_{\alpha,\beta}\notag \\
&=[C_X(k_y)]_{(L_x+1-x)\alpha,(L_x+1-x')\beta}.
\end{align}
Writing this equation in a matrix form (and suppressing explicitly writing $k_y$ dependence), it looks
\begin{align}
    C_X^* = 
    \begin{pmatrix}
        \mathbf{0} & \mathcal{P}_0 \\ \mathcal{P}_0 & \mathbf{0}
    \end{pmatrix}
    C_X
    \begin{pmatrix}
        \mathbf{0} & \mathcal{P}_0 \\ \mathcal{P}_0 & \mathbf{0}
    \end{pmatrix},
\end{align}
where
\begin{align}
    \mathcal{P}_0&=
\begin{pmatrix}
&&\mathbf{1}_u\\
&\reflectbox{\text{$\ddots$}}&\\
\mathbf{1}_u&&
\end{pmatrix},
\end{align}
where $\mathbf{1}_u$ is a unit matrix in the unit cell. 
Looking at each block, we find the following relations
\begin{align}
C_{AA}&=\mathcal{P}_0 C_{AA}^{*}\mathcal{P}_0,
\\
C_{BA}\mathcal{P}_0&=\mathcal{P}_0 C_{BA}^{*}.
\label{CRL_P}
\end{align}
Therefore, if we have an eigenstate $\ket{\psi_0}$ of $C_{AA}$, $C_{AA} \ket{\psi_0} = \xi \ket{\psi_0}$, we can obtain another eigenstate of $C_{AA}$, by $\mathcal{P}_0\ket{\psi_0}^{*}$ with the same eigenvalue because
\begin{align}
    C_{AA} \mathcal{P}_0 |\psi_0\rangle^* &= 
    \mathcal{P}_0 C_{AA}^{*}\mathcal{P}_0 \mathcal{P}_0 |\psi_0\rangle^*
    \notag \\
    &=
    \mathcal{P}_0 \left( C_{AA} |\psi_0\rangle \right)^*
    \notag \\&=
    \xi \mathcal{P}_0 |\psi_0\rangle^*.
\end{align}
Note that since $\mathcal{P}_0$ flips the position in $x$ direction, if $|\psi_0\rangle$ is localized on the right edge, $\mathcal{P}_0|\psi_0\rangle^*$ is localized on the left edge.

We can also define an operation $\Xi$ that change both the eigenvalues and on which side of the edges the eigenstates are localized, by using $C_{BA}$\cite{PhysRevB.83.245132,PhysRevB.85.165120}. (We note that we do not need to require $PT$ symmetry to have $\Xi$; the existence of $\Xi$ is a general property of the correlation matrix.)
Using the property $C_X^{2}(k_y)=C_X(k_y)$, the following relations hold:
\begin{align}
C^{\dagger}_{BA}(k_y)C_{BA}(k_y)&=C_{AA}(k_y)(1-C_{AA}(k_y)),
\label{CRLd_CRL}
\\
C_{AA}(k_y)C_{BA}(k_y)&=C_{BA}(k_y)(1-C_{AA}(k_y)).
\label{CR_CRL}
\end{align}
For an eigenstate $|\psi_i\rangle$ of $C_{AA}(k_y)$ with an eigenvalue $\xi_i(k_y)$, $C_{AA}(k_y)|\psi_i\rangle = \xi_i (k_y)|\psi_i\rangle$, the following relation holds from Eq.(\ref{CR_CRL}):
\begin{align}
C_{AA}(k_y)C_{BA}(k_y)\ket{\psi_i}
&=C_{BA}(k_y)(1-C_{AA}(k_y))\ket{\psi_i}
\notag \\
&=(1-\xi_i(k_y))C_{BA}(k_y)\ket{\psi_i}.
\end{align}
The norm of $C_{BA}(k_y)\ket{\psi_i}$ is calculated by using Eq.(\ref{CRLd_CRL}):
\begin{align}
||C_{BA}(k_y)\ket{\psi_i}||^2=\xi_i(k_y)(1-\xi_i(k_y)).
\end{align}
Therefore, for $\xi_i(k_y)\neq0,1$ we obtain the eigenstate $\Xi\ket{\psi_i}$ of $C_{AA}(k_y)$ with the eigenvalue $(1-\xi_i(k_y))$:
\begin{align}
\Xi\ket{\psi_i}\equiv\frac{C_{BA}(k_y)}{\sqrt{\xi_i(k_y)(1-\xi_i(k_y))}}\ket{\psi_i}. 
\end{align}
The operator $\Xi$ can be expressed as
\begin{align}
    \Xi=\sum_{i}\frac{C_{BA}(k_y)}{\sqrt{\xi_i(k_y)(1-\xi_i(k_y))}}\ket{\psi_i}\bra{\psi_i}. \label{Xi_xi_psi}
\end{align}
In particular, noting that $\epsilon_i=1/2-\xi_i$ from Eq.~(\ref{H_flat_CL}), Eq.~(\ref{Xi_xi_psi}) is equivalent to Eq.~(5). 
Furthermore, for an eigenstate $\ket{\psi_i}$ with an eigenvalue $\epsilon_i$ of $H_{\text{flat}}^{x\text{-OBC}}(k_y)$, $\Xi\ket{\psi}$ has an eigenvalue $\epsilon_i^{\prime}=(1/2-(1-\xi_i))=-\epsilon_i$.

We note that, $C_{BA}(k_y)$ as a matrix has nonzero components typically at lower left and upper right regions. As a result, $\Xi$ changes the side of localization of eigenstates; if $|\psi_i\rangle$ is localized at the right edge of $A$ region, $\Xi|\psi_i\rangle$ is localized at the left edge of $A$ region.

Combining the two operations of $PT$ and $\Xi$, we get the following anti-linear operation $(PT)\Xi$:
\begin{align}
(PT)\Xi
&\equiv \sum_i \frac{\mathcal{P}_0 C_{BA}^{*}(k_y)}{\sqrt{\xi_i(k_y)(1-\xi_i(k_y))}}\Big{(}\ket{\psi_i}\bra{\psi_i}\Big{)}^{*}K. 
\end{align}
Note that this operation is defined only if the $C_L$ eigenvalue $\xi_i$ of $\ket{\psi_i}$ is neither $0$ nor $1$. The $C_L$ eigenvalue of $(PT)\Xi\ket{\psi_i}$ is $(1-\xi_i)$. 

We prove that $(PT)\Xi$ is an anti-unitary operator and squares to $+1$. We again take $\ket{\psi_i}$ to be an eigenstate of $C_{AA}$ with an eigenvalue $\xi_i$, and $\ket{\widetilde{\psi_i}}:=(PT)\Xi\ket{\psi_i}$. First, $(PT)\Xi$ preserves the norm:
\begin{align}
&\langle\widetilde{\psi_i}\ket{\widetilde{\psi_i}}\notag \\
=&\frac{1}{\xi_i(k_y)(1-\xi_i(k_y))}(\bra{\psi_i})^{*}C_{BA}^{T}(k_y)\mathcal{P}_0^{T}\mathcal{P}_0 C_{BA}^{*}(k_y)\ket{\psi_i}^{*}\notag \\
=&\frac{1}{\xi_i(k_y)(1-\xi_i(k_y))}\big{[}\bra{\psi_i}C_{AA}(k_y)(1-C_{AA}(k_y))\ket{\psi_i}\big{]}^{*}\notag \\
=&\langle \psi_i\ket{\psi_i}.
\end{align}
Here in the second line, we used Eq.~(\ref{CRLd_CRL}). 
Also, $(PT)\Xi$ preserves orthogonality:
\begin{align}
&\langle\widetilde{\psi_i}\ket{\widetilde{\psi_j}}\cdot\sqrt{\xi_i(k_y)(1-\xi_i(k_y))}\sqrt{\xi_j(k_y)(1-\xi_j(k_y))}
\notag \\
=&\big{[}\bra{\psi_i}C_{AA}(k_y)(1-C_{AA}(k_y))\ket{\psi_j}\big{]}^{*}\notag \\
=&\langle \psi_j\ket{\psi_i}=0\quad(\xi_i\neq \xi_j).
\end{align}
Therefore, $(PT)\Xi$ is an anti-unitary operator.
Finally, we show that $(PT)\Xi$ squares to 1:
\begin{align}
&(PT)\Xi\ket{\widetilde{\psi_i}}
\notag \\
=&\frac{1}{\xi_i(k_y)(1-\xi_i(k_y))}\mathcal{P}_0 C_{BA}^{*}(k_y)\Big{(}\mathcal{P}_0 C_{BA}^{*}(k_y)\ket{\psi_i}^{*}\Big{)}^{*}
\notag \\
=&\frac{1}{\xi_i(k_y)(1-\xi_i(k_y))}C_{BA}^{\dagger}(k_y)\mathcal{P}_0 \mathcal{P}_0^{*} C_{BA}(k_y)\ket{\psi_i}
\notag \\
=&\frac{1}{\xi_i(k_y)(1-\xi_i(k_y))}C_{BA}^{\dagger}(k_y)C_{BA}(k_y)\ket{\psi_i}
\notag \\
=&\frac{1}{\xi_i(k_y)(1-\xi_i(k_y))}C_{AA}(k_y)\left( 1 - C_{AA}(k_y)\right)\ket{\psi_i} \notag \\
=&\ket{\psi_i}.
\end{align}
Since both $PT$ and $\Xi$ changes the side of localization, the combined operation $(PT)\Xi$ thus transforms one state localized at one edge to another state with an opposite energy localized at the same edge.

\section{Wilson loop spectra}\label{Appendix B}
\begin{figure}[t]
\centerline{\includegraphics[width=8.5cm,clip]{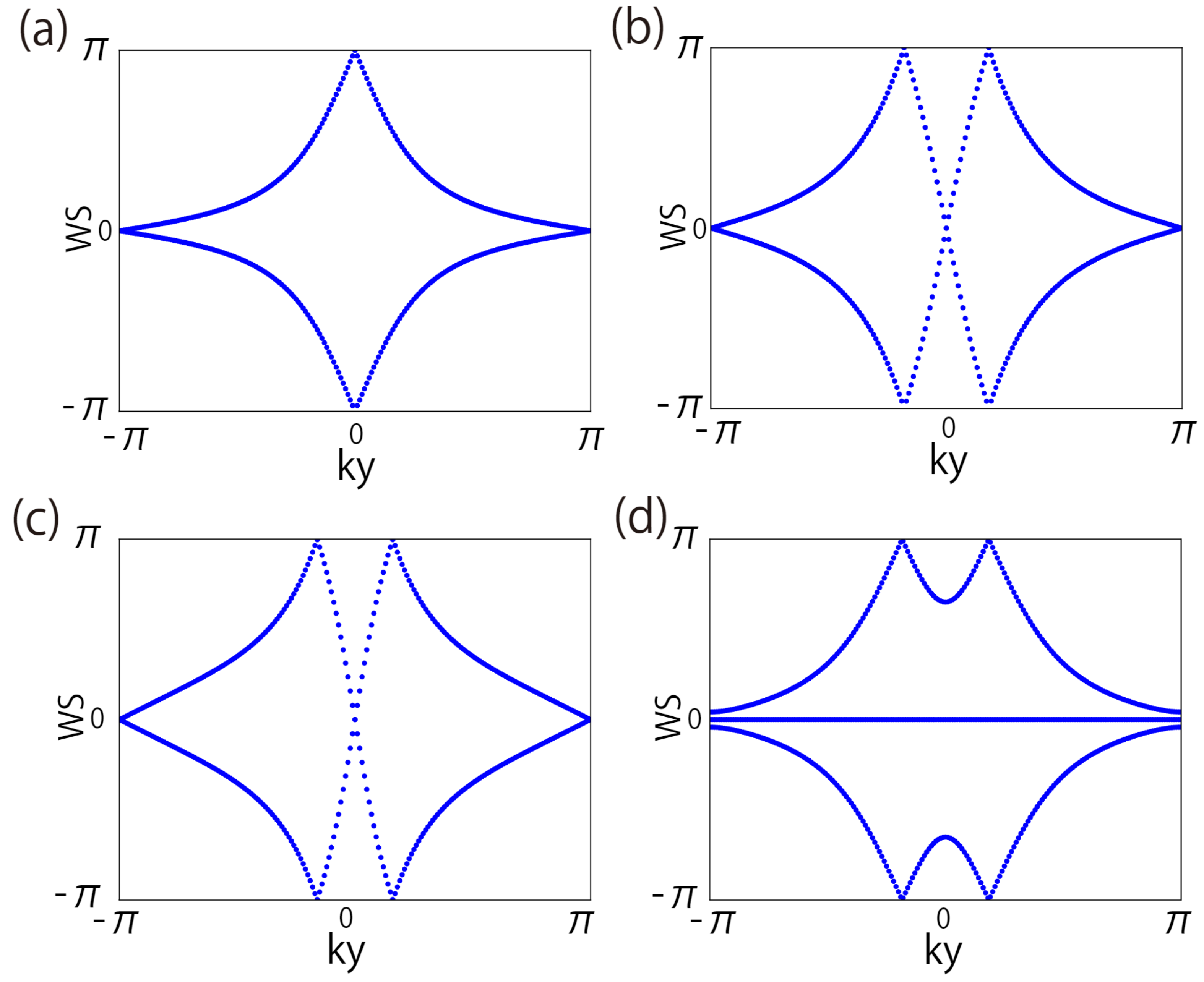}}
  \caption{Wilson loop spectra for (a) $H_{\text{SW}}$, (b) $H_{\text{Euler}}$ and that for $H_{\text{Euler+NI}}$ with (c) $V_0=1.5$ and (d) $V_0=-0.5$. By counting the Wilson loop windings, we find (a) $\chi=1$, (b) $\chi=2$, and (c) $\chi=2$, respectively. In (d), $w_2$ is 0 from the number of $\pi$ crossings in the Wilson loop spectra.
  }
    \label{EZ_WL}
\end{figure}

The Euler class topology can be detected from the Wilson loop winding\cite{PhysRevLett.121.106403,PhysRevB.102.115135,bouhon2020non}.
Specifically, Wilson loop winding gives the absolute value of the Euler class $|\chi|$\cite{PhysRevB.102.115135}. 
Since the sign of the Euler class is ambiguous, unless otherwise stated, we assume that the Euler class $\chi$ is positive by taking an appropriate basis\cite{bouhon2020non}. 
Figure \ref{EZ_WL} shows Wilson loop spectra for several models used in the main text. 
Figures \ref{EZ_WL}(a-c) show that they all have non-trivial Euler classes.

The case of Fig.~\ref{EZ_WL}(d) with three occupied bands is characterized by the second Stiefel-Whitney class $w_2$ instead of the Euler class. From the number of $\pi$-crossing Wilson loop spectra, $w_2=0$ (mod $2$) in this case\cite{PhysRevLett.121.106403}.

\section{Effective Hamiltonian}\label{Appendix C}
Since $(PT)\Xi$ is an antiunitary operation that squares to +1, we can take a basis in which $(PT)\Xi$ is expressed as the complex conjugate $K$. 
(Note that, in this new basis, $PT$ is no longer expressed by $K$.)
Here, we briefly explain how we can obtain a basis such that $(PT)\Xi=K$. 
From a pair of eigenstates of $H_\text{flat}^{x\text{-OBC}}$ localized on one edge, $|\psi_j\rangle$ and $(PT)\Xi |\psi_j\rangle$, we construct the following two states:
\begin{align}
\ket{\psi_{+}^{(j)}}&:=\frac{1}{\sqrt{2}}\big{(}\ket{\psi_j}+(PT)\Xi\ket{\psi_j}\big{)},
\\
\ket{\psi_{-}^{(j)}}&:=\frac{1}{\sqrt{2}i}\big{(}\ket{\psi_j}-(PT)\Xi\ket{\psi_j}\big{)}.
\end{align}
These states are invariant under $(PT)\Xi$, that is, $(PT)\Xi \ket{\psi_{+}^{(j)}} = \ket{\psi_{+}^{(j)}}$ and $(PT)\Xi \ket{\psi_{-}^{(j)}} = \ket{\psi_{-}^{(j)}}$. This invariance combined with the fact that $(PT)\Xi$ is an anti-unitary operator implies that, in the space spanned by $\ket{\psi_{+}^{(j)}}$ and $\ket{\psi_{-}^{(j)}}$, $(PT)\Xi$ acts as a simple complex conjugation $K$.
Taking all the pairs of eigenstates we consider to build an effective Hamiltonian and considering the following basis
\begin{align}
\bigg{[}\ket{\psi_{+}^{(1)}},\ket{\psi_{-}^{(1)}},\ket{\psi_{+}^{(2)}},\cdots,\ket{\psi_{-}^{(N_0)}}\bigg{]},
\end{align}
where $N_0$ is the number of pairs of eigenstates, or half the number of all the eigenstates, one wishes to include in the effective Hamiltonian.
In this basis, $(PT)\Xi$ acts as $K$. Since the effective Hamiltonian should anti-commute with $(PT)\Xi = K$, the effective Hamiltonian should be a purely imaginary matrix, and thus it should be a skew-symmetric matrix of the following form:
\begin{align}
H_{\text{eff}}^{x\text{-OBC}}=
\begin{pmatrix}
\epsilon_1\sigma_y&&\\
&\ddots&\\
&&\epsilon_{N_0}\sigma_y
\end{pmatrix},
\label{HL_skew}
\end{align}
which is the desired form.

\section{Analysis of more models}\label{Appendix D}
Here, we perform numerical calculations for more models to give further support to our conjecture.

\subsection{Kobayashi-Furusaki model}
\begin{figure}
\centerline{\includegraphics[width=8.5cm,clip]{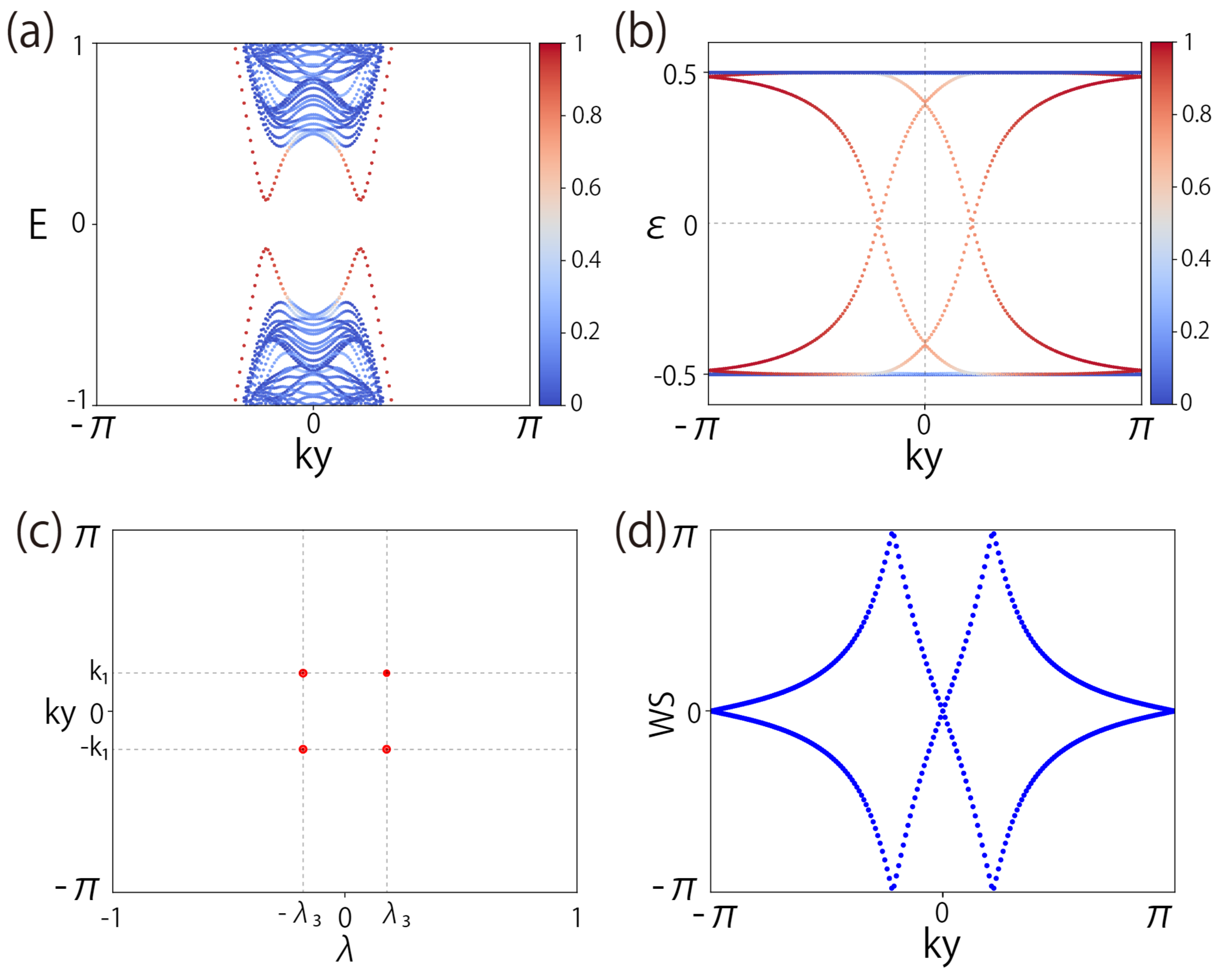}}
  \caption{Numerical calculation for the model $H_{S1}$. (a) Edge spectrum.  Colors represent the squared amplitudes of the states in the edge unit cell. (b) Entanglement spectrum. (c) Spectral flow with cutting procedure. Here, $\lambda_3=0.18$ and $k_1=0.66$.  (d) Wilson loop spectrum. 
  }
    \label{Kobayashi}
\end{figure}
First, we use the model introduced by Kobayashi and Furusaki in the study of fragile topological insulators \cite{PhysRevB.104.195114}. The restriction of their model to $k_z=0$ corresponds to an insulator with Euler class $|\chi|=2$: 
\begin{align}
H_{S1}(\bm{k})
=&\Bigg{(}3-2\sum_{i=x,y}\cos(k_i)\Bigg{)}\tau_0\sigma_z
\notag \\
&+\bigg{(}2(\cos(k_x)-\cos(k_y))+\mathcal{M}_1\bigg{)}\tau_z\sigma_x
\notag \\
&+\Big{(}2\sin(k_x)\sin(k_y)+\mathcal{M}_2
\Big{)}\tau_x\sigma_x
\notag \\
&+\Delta\tau_x\sigma_0,
\label{Kobayashi_H}
\end{align}
where $\mathcal{M}_1=0.1$, $\mathcal{M}_2=0.08$ and $\Delta=0.2$. The $PT$ operation is represented as complex conjugation $K$.  Here, the term $\Delta\tau_x\sigma_0$, which is absent in~\cite{PhysRevB.104.195114}, is added to open a gap in the edge spectrum as shown in Fig.~\ref{Kobayashi}(a). The Wilson loop spectrum (Fig.~\ref{Kobayashi}(d)) confirms that the model has Euler class $|\chi|=2$.
Unlike the edge spectrum, the entanglement spectrum is gapless as shown in Fig.~\ref{Kobayashi}(b). There are two linear crossings at $\epsilon=0$, consistent with our conjecture for $|\chi| = 2$. Furthermore, from the spectral flow of the cutting procedure (see Fig.~\ref{Kobayashi}(c)), we can see that there are four crossings as we anticipate.

\subsection{$C_6$-symmetric models}
Next, we confirm that our results also hold for systems other than square lattices. As examples, we consider two models of $C_6$ symmetric lattices.

\begin{figure}[t]
\centerline{\includegraphics[width=8.5cm,clip]{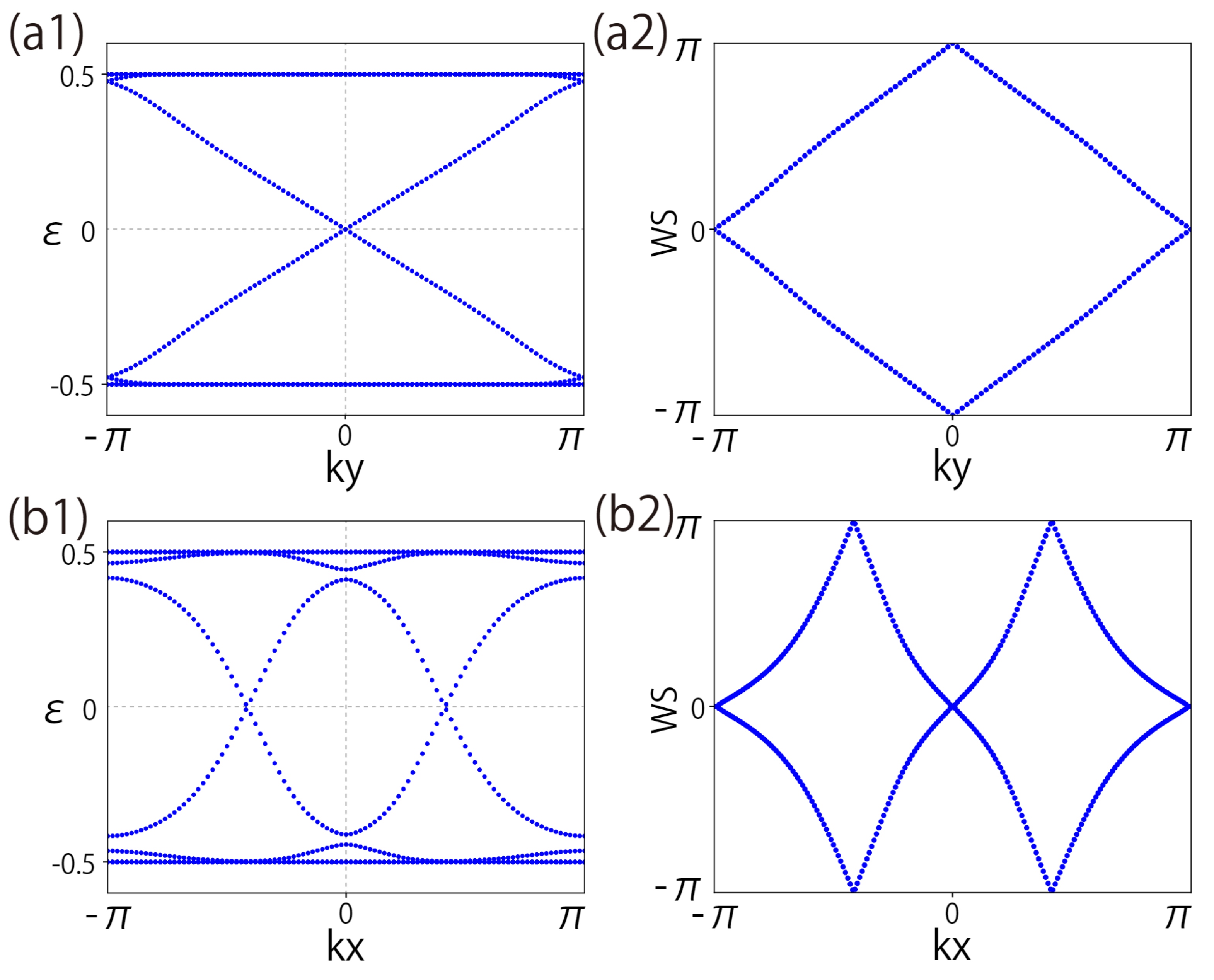}}
  \caption{Numerical calculations for the models (a) $H_{S2}$ and (b) $H_{S3}$. (a1,b1) Entanglement spectra, and (a2,b2) Wilson loop spectra. 
  }
    \label{C6_ES_WL}
\end{figure}

The following honeycomb lattice model is an effective model for the nearly flat bands in twisted bilayer graphene at the magic angle\cite{PhysRevB.98.085435}:
\begin{align}
H_{S2}(\bm{k})
=&\Bigg{(}1+2\cos\frac{\sqrt{3}k_x}{2}\cos\frac{k_y}{2}\Bigg{)}(0.4\sigma_x+0.6\tau_z\sigma_x)
\notag \\
&+2\sin\frac{\sqrt{3}k_x}{2}\cos\frac{k_y}{2}(0.4\sigma_y+0.6\tau_z\sigma_y)
\notag \\
&+\bigg{(}4\cos\frac{\sqrt{3}k_x}{2}\sin\frac{k_y}{2}-2\sin(k_y)\bigg{)}(0.1\tau_x). 
\end{align}
Here, the $PT$ operation is represented as $\sigma_x K$. 
Figures~\ref{C6_ES_WL}(a1,a2) show the entanglement spectrum and Wilson loop spectrum of $H_{S2}(\bf{k})$. The Wilson loop winding in [(a2)] shows that the Euler number is $|\chi| = 1$.
The entanglement spectrum [(a1)] shows one linear crossing point at $\epsilon=0$, which is in agreement with our conjecture with $|\chi|=1$.

The final model is on a triangular lattice, originally used in the study of triple nodal points in~\cite{PhysRevB.106.085129}. Restricting the model to $k_z=0$ results in the following model:
\begin{align}
&\quad\ H_{S3}(\bm{k})
\notag \\
&=-\Big{[}t_1+t_2\Big{(}\cos k_x+2\cos\frac{k_x}{2}\cos\frac{\sqrt{3}k_y}{2}\Big{)}+t_3\Big{]}\tau_z\sigma_0
\notag \\
&\quad\ -t_4(-\tau_z+\tau_z\sigma_z)
\notag \\
&\quad\ -t_5\Big{[}\Big{(}\cos k_x-\cos\frac{k_x}{2}\cos\frac{\sqrt{3}k_y}{2}\Big{)}(-\tau_x\sigma_z-\tau_x)
\notag \\
&\quad\ +\sqrt{3}\sin\frac{k_x}{2}\sin\frac{\sqrt{3}k_y}{2}(-\tau_z-\tau_z\sigma_z)\Big{]}
\notag \\
&\quad\ -\sqrt{2}t_6\Big{[}\Big{(}\cos k_x-\cos\frac{k_x}{2}\cos\frac{\sqrt{3}k_y}{2}\Big{)}(\tau_x\sigma_x-\tau_z\sigma_x)
\notag \\
&\quad\ -\sqrt{3}\sin\frac{k_x}{2}\sin\frac{\sqrt{3}k_y}{2}(\tau_x\sigma_x+\tau_z\sigma_x)\Big{]},
\end{align}
where $t_1=4$, $t_2=-\frac{2}{3}$, $t_3=-3$, $t_4=-\frac{1}{2}$, $t_5=-\frac{1}{3}$ and $t_6=\frac{6}{5}$. 
Here, the $PT$ operation is represented as complex conjugation $K$. 
Figures~\ref{C6_ES_WL}(b1,b2) show the entanglement spectrum and Wilson loop spectrum of $H_{S3}(\bf{k})$. The Wilson loop spectrum shows that the Euler class is $|\chi| = 2$. The entanglement spectrum [(b1)] has two linear crossing points at $\epsilon=0$, which is consistent with our conjecture with $|\chi|=2$.

\end{document}